\documentclass[10pt,conference]{IEEEtran}

\usepackage{amsmath,amssymb,amsfonts}
\usepackage{algorithmic}
\usepackage{textcomp}
\usepackage{xcolor}
\def\BibTeX{{\rm B\kern-.05em{\sc i\kern-.025em b}\kern-.08em
    T\kern-.1667em\lower.7ex\hbox{E}\kern-.125emX}}

\usepackage[utf8]{inputenc}
\usepackage{graphicx}
\usepackage{wrapfig}
\graphicspath{ {./images/} }
\usepackage[english]{babel}
\usepackage[sorting=none]{biblatex}
\usepackage[a4paper, margin=1in]{geometry}

\usepackage{xcolor}

\usepackage{hyperref}
\usepackage{bm}
\usepackage{qcircuit}
\usepackage{braket}
\usepackage{amsmath}
\usepackage{amssymb}

\title{\huge{Adapting Quantum Approximation Optimization Algorithm (QAOA) for Unit Commitment}
}

\author{
\IEEEauthorblockN{
Samantha Koretsky\IEEEauthorrefmark{1}, Pranav Gokhale\IEEEauthorrefmark{2}, Jonathan M. Baker\IEEEauthorrefmark{1}, Joshua Viszlai\IEEEauthorrefmark{1}, Honghao Zheng\IEEEauthorrefmark{3},\\Niroj Gurung\IEEEauthorrefmark{3}, Ryan Burg\IEEEauthorrefmark{3}, Esa Aleksi Paaso\IEEEauthorrefmark{3}, Amin Khodaei\IEEEauthorrefmark{4}, Rozhin Eskandarpour\IEEEauthorrefmark{5}, Frederic T. Chong\IEEEauthorrefmark{1}\IEEEauthorrefmark{2}}

\IEEEauthorblockA{\IEEEauthorrefmark{1}University of Chicago} \IEEEauthorblockA{\IEEEauthorrefmark{2}Super.tech} \IEEEauthorblockA{\IEEEauthorrefmark{3}Commonwealth Edison}
\IEEEauthorblockA{\IEEEauthorrefmark{4}University of Denver}
\IEEEauthorblockA{\IEEEauthorrefmark{5}Resilient Entanglement}
}

\begin{document}

\maketitle
\thispagestyle{plain}
\pagestyle{plain}

\begin{abstract}
    In the present Noisy Intermediate-Scale Quantum (NISQ), hybrid algorithms that leverage classical resources to reduce quantum costs are particularly appealing. We formulate and apply such a hybrid quantum-classical algorithm to a power system optimization problem called Unit Commitment, which aims to satisfy a target power load at minimal cost. Our algorithm extends the Quantum Approximation Optimization Algorithm (QAOA) with a classical minimizer in order to support mixed binary optimization. Using Qiskit, we simulate results for sample systems to validate the effectiveness of our approach. We also compare to purely classical methods. Our results indicate that classical solvers are effective for our simulated Unit Commitment instances with fewer than 400 power generation units. However, for larger problem instances, the classical solvers either scale exponentially in runtime or must resort to coarse approximations. Potential quantum advantage would require problem instances at this scale, with several hundred units.
\end{abstract}


\section{Background}
Quantum computing has the potential to revolutionize computation in specific domains. Google’s recent quantum supremacy experiment demonstrated that quantum computers have the potential to perform computations that outclass the world's fastest supercomputers \cite{arute2019quantum}. Though still in early stages, quantum capabilities are advancing rapidly by providing novel ways to solve previously intractable problems. In this paper, we aim to apply quantum computing to achieve a practical advantage over current methods in a power systems application.

Current quantum machines are limited in that they are noisy and only accommodate a small number of qubits, generally under 100. Additionally, gates are prone to error. For this reason, current approaches favor hybrid algorithms that combine classical and quantum methods to minimize required qubit and gate counts. We apply this hybrid architecture to smart grid optimization, specifically for the unit commitment problem. 
    
\subsection{Unit Commitment}
Unit Commitment (UC) is an important optimization problem in the electrical power industry. It aims to minimize operational cost while meeting a target power load using a number of power-generating units that are subject to constraints \cite{ajagekar2019quantum}. As the size and complexity of current energy systems increase, improving efficiency of solving UC becomes more important to industry. 

\begin{figure*}[h!]
    \centering
    \includegraphics[width=0.85\textwidth]{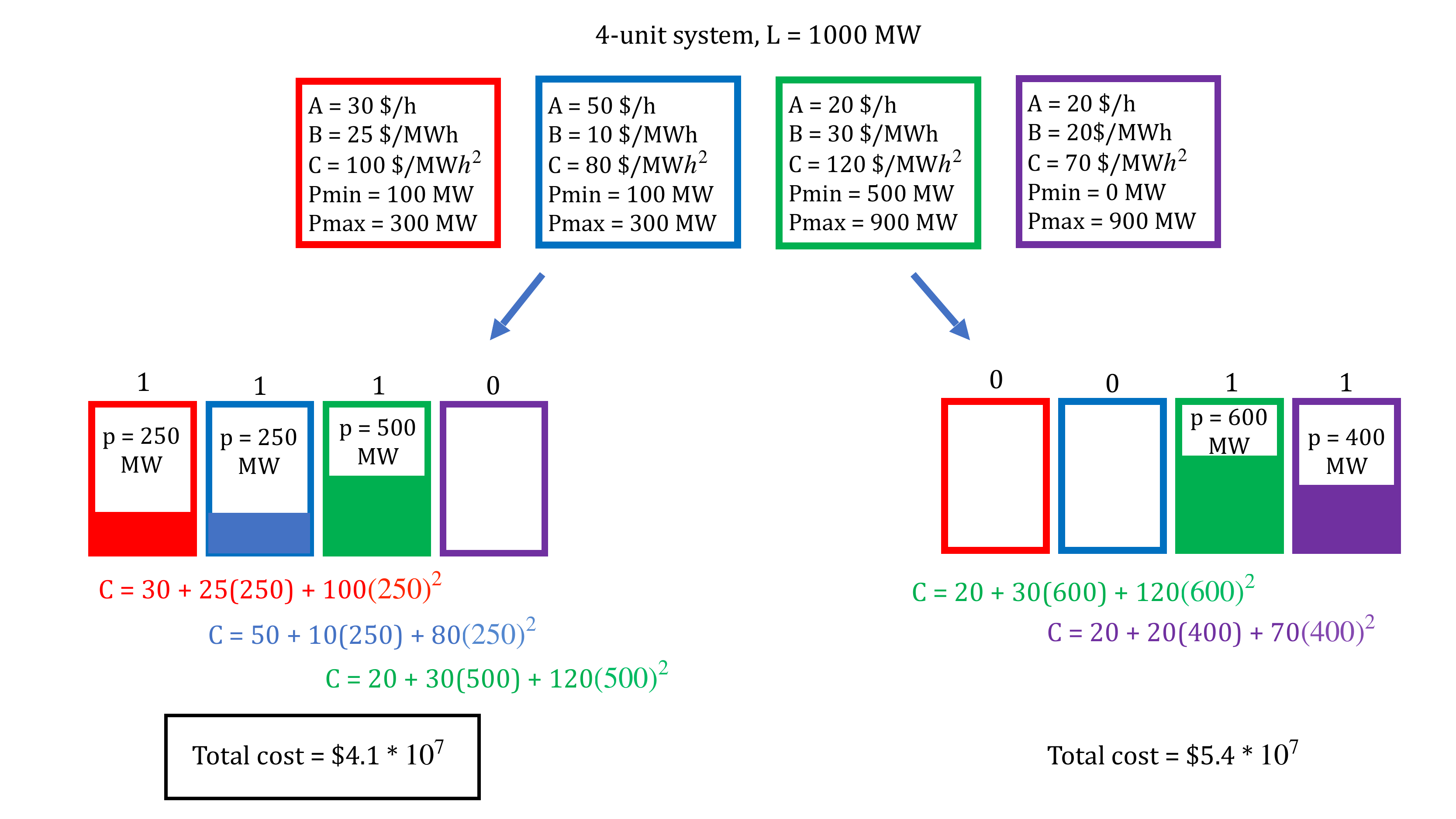}
    \caption{A diagram of two ways that units in a 4-unit system could meet a given power load, subject to the constraints of the units, in one time step. The total cost of each configuration is given by the sum of the cost of each unit that is turned on. Out of these two systems, the left is more cost efficient. More details on how the cost of a unit is calculated will be given in Sec. \ref{sec:our_approach}.}
    \label{fig:uc}
\end{figure*}

In UC, the main goal is to meet the power load $L$ while minimizing the total cost. Each unit's cost function is defined by three coefficients: $A$ is the fixed cost coefficient that a unit necessarily incurs when it is turned on, regardless of the power it contributes. $B$ and $C$ are the linear and quadratic coefficients, respectively, and contribute to the unit's cost based on its power level. 
Each unit is further subject to a set of operational constraints, including but not limited to minimum and maximum generation limits, ramping up and down limits, minimum on and off time constraints, and reserve. In this paper, we only focus on minimum and maximum power generation limits, $p_{min}$ and $p_{max}$. 
Additionally, individual power units may be turned on or off, adding another layer of complexity to the problem. 
The total cost is given by the sum of the costs only of the units that are on. 

UC aims to determine the optimal combination of units to use and the power levels at which they should operate, all while minimizing the total cost, meeting the load $L$, and following the constraints defined by the coefficients $p_{min}$ and $p_{max}$ for each unit. 
Fig.~\ref{fig:uc} demonstrates two ways that the load of a 4-unit system could be met while keeping within each unit's constraints, and gives the cost of each configuration.
    
\subsection{Quantum Approximation Optimization Algorithm}
The Quantum Approximation Optimization Algorithm (QAOA) \cite{farhi2014quantum} is a quantum algorithm that can be used to approximate solutions to optimization problems. 
It is a hybrid algorithm that uses both quantum and classical resources with the aim of reducing the resource requirements on the quantum computer relative to a quantum-only algorithm.
QAOA is designed for quadratic unconstrained binary optimization (QUBO) problems, which are a specific kind of combinatorial optimization problem that are somewhat similar to UC. Both problems feature binary choices, although QUBO is defined by only discrete variables, while UC contains both continuous and discrete variables. The similarities between QUBO and UC make QAOA a compelling algorithm to use in solving UC. 

The job of the classical computer in QAOA is to optimize a set of \textit{variational parameters}: $\bm{\gamma} = (\gamma_1, \gamma_2, ..., \gamma_P)$ and $\bm{\beta} = (\beta_1, \beta_2, ..., \beta_P)$ where $\gamma_i, \beta_i \in [0, 2\pi)$. A quantum computer will execute a circuit which is a function of these vectors of angles $\bm{\gamma}$ and $\bm{\beta}$. The length of parameter vectors, $P$, is proportional to the depth (runtime) of the QAOA circuit and does not depend on the total number of qubits. Initially, the quantum computer puts the $N$ qubits into a uniform superposition over all $2^N$ bitstrings. Next, the quantum computer executes a two-step sequence. First a \textit{cost Hamiltonian} is applied that phases\footnote{This is a uniquely quantum phenomenon that enables destructive interference when summing bitstrings.} each bitstring by a quantity related to that bitstring's cost and to $\gamma_1$. Second, a \textit{mixing Hamiltonian} is applied so that probability amplitude can transfer between the $2^N$ bitstrings. This two-step sequence---cost Hamiltonian parametrized by $\gamma_i$ and mixing Hamiltonian parametrized by $\beta_i$---is repeated for $i = 2, 3, ..., P$.

By sampling from the output of the quantum circuit with different $\bm{\gamma}, \bm{\beta}$ vectors, a score can be recorded for each choice of variational parameters. The classical computer then performs an outer loop (with feedback) to optimize over the $2P$-dimensional space, with the aim of producing the optimal parameters: $\bm{\gamma^*}, \bm{\beta^*}$. When the quantum computer is evaluated with these optimal (or near-optimal parameters), it will output bitstrings that approximately extremize the QUBO objective function.

At $P$ approaches $\infty$, QAOA can recover the Quantum Adiabatic Algorithm, which would \textit{exactly} solve the target optimization problem \cite{farhi2001quantum}. However, this would also take infinite circuit depth. Instead, we are interested in the performance of QAOA at small, finite $P$. Choices such as $P=1$ or $P=2$ are particularly favorable because they require low gate count and circuit depth. This is suitable for near-term quantum hardware since (a) gates are significant error rates and (b) qubit lifetimes are short, which prevents deep circuits from running effectively.

\section{Prior Work}
There are a variety of classical approaches to solving or approximating Unit Commitment instances. For example, Generic Algebraic Modeling System (GAMS) can solve mixed-binary optimization programming problems like UC through tools like DICOPT a DIscrete and Continuous OPTimizer \cite{gams}. As in our quantum approach presented in Section~\ref{sec:our_approach}, DICOPT relaxes equality constraints and penalizing violations of constraints using slack variables \cite{gams}. Thus, the program ultimately approximates a solution to a given Unit Commitment instance. Other classical solvers include IBM's CPLEX which is a popular choice for both exactly and approximately solving MBO problems in addition to Mixed-Integer Programming (MIP) problems and other more general linear programming problems \cite{cplex}. CPLEX makes use of a branch-and-bound approach to solving these problems classically.

To the best of our knowledge, \cite{ajagekar2019quantum} is the only prior work on solving Unit Commitment with a quantum approach. The authors propose an algorithm based on quantum annealing. To cope with the mixture of continuous and binary variables, \cite{ajagekar2019quantum} opts to discretize the continuous variables with a one-hot encoding. As a result, $N(\ell+2)$ qubits are required to solve an $N$-unit system discretized to have $\ell$ partitions between $p_{min,i}$ and $p_{max,i}$. While the results are promising at small-scale, the one-hot encoding of continuous variables is expensive in qubit count. Our approach aims to avoid this discretization cost. In addition, we use QAOA (instead of the quantum adiabatic algorithm), which enables our approach to be executed on gate-model quantum computers.

Finally, three recent papers have proposed quantum algorithms for mixed binary optimization problems more broadly. \cite{braine2019quantum} uses QAOA as a lever for translating the mixed binary optimization problem into a continuous optimization problem. The approach is demonstrated for transaction settlement, a financial application. Though developed independently, our approach in \ref{sec:our_approach} is similar to \cite{braine2019quantum}, but applied to Unit Commitment. \cite{gambella2020multiblock} takes another approach, by using the Alternating Direction Method of Multipliers (ADMM) to convert an input mixed binary optimization problem into a collection of QUBOs. Finally, \cite{chang2020quantum} applies Bender's decomposition \cite{bnnobrs1962partitioning} to divide an input mixed-integer linear program into a collection of quantumly-solvable QUBOs and classically-solvable linear programs.

\section{Our QAOA-based Approach} \label{sec:our_approach}
The critical insight behind our work is that QAOA converts discrete optimization problems (e.g. over bitstrings) into a continuous optimization problem (over the variational parameters $\bm{\gamma}, \bm{\beta}$). Our approach to solving UC combines quantum and classical methods. 
We use QAOA to handle the binary variables of the problem (whether a unit is on or off) and a classical optimizer to handle the continuous variables (how much power each unit should provide). The classical optimizer comprises the outer loop of our algorithm, and QAOA the inner. Thus, QAOA never needs to discretize continuous variables into qubits. Meanwhile, in the frame of reference of the classical optimizer, the mixed integer problem becomes a continuous one. 
This combination of classical and quantum algorithms approximates a solution while both limiting the number of gates used in the quantum computation and allowing for a plausible quantum advantage. 

\subsection{Formulation}
In the Unit Commitment problem we have $N$ units. Each unit may be turned on or off, given by a binary variable $y_i \in \set{0, 1}$. When $y_i = 1$, the power of the unit $p_i$ is a continuous real value constrained as $p_{min, i} \le p_i \le p_{max, i}$. When $y_i = 0$, $p_i = 0$. Together these restrictions gives a quadratic constraint: $p_{min, i}y_i \le p_i \le p_{max, i}y_i$. UC aims to turn on certain units (decide which $y_i$ should be 1) and set their corresponding power values (decide the value of $p_i$) so that the sum $p_i$ is exactly $L$. 

Each power unit has a corresponding function $H(y_i, p_i)$ which specifies the cost of turning on unit $i$ and for generating a certain amount of power $p_i$. This cost function is quadratic for unit commitment:

\begin{equation}
    H(y_i, p_i) = A_iy_i + B_ip_i + C_ip_i^2
\end{equation}

where $A_i, B_i, C_i \in \mathbb{R}$ are constant. This overarching objective of a UC problem then is to minimize the sum of these cost functions. The following equations summarize a generic UC problem.

\begin{align}
\label{UC-spec}
\text{min} \quad \sum_{i = 1}^{N} H(y_i, p_i) &  \\
\text{s.t.} \quad \sum_{i=1}^N p_i &= L \\
p_iy_i &\le p_{max, i}, & p_iy_i &\ge p_{min, i} \\
p_i &\in \mathbb{R} & y_i &\in \set{0, 1}
\end{align}

We convert this formulation into a QUBO problem. The goal is to convert the constrained variables $y_i, p_i$ to become unconstrained by modifying the objective function to include penalty terms so that when the values $y_i, p_i$ are outside of their desired ranges the objective suffers. Now we let $p_i \ge 0$. For each continuous variable $p_i$ we introduce two slack variables $s_{i, 1}$, $s_{i, 2} \in \mathbb{R}_{\ge 0}$, the first to penalize if $p_i$ is less than the minimum and the second to penalize if $p_i$ is larger than its maximum. Consider the following penalty term

\begin{equation}
    \lambda(p_i - s_{i, 1} - p_{min, i}y_i)^2
\end{equation}
where $\lambda$ is some fixed constant which must be set empirically. Suppose $y_i = 1$ and $p_i$ is greater than $p_{min, i}$ then this equation can be minimized by setting $s_{i, 1} = p_i - p_{min, i} > 0$ resulting in 0 penalty. However, if $p_i < p_{min, i}$ then this equation is minimized when $s_{i, 1} = 0$ resulting in a positive penalty term. This equation effective encodes the previous constraint bounding $p_i \ge p_{min, i}$ by guiding the objective into a solution which minimizes this penalty term. We can do the same by adding in a penalty term using $s_{i, 2}$ to penalize values of $p_i > p_{max, i}$. Finally, we need a penalty term to encode the constrain that the sum of the $p_i$ is the power load $L$. The following term
\begin{equation}
    \lambda\big( \sum_{i = 1}^N p_iy_i - L\big)^2
\end{equation}
is minimized when the sum is exactly $L$. In practice, we may want to choose different $\lambda$ values for each penalty term to enforce certain constraints more or less. Putting it all together we obtain the following objective function which ideally encodes the same UC problem
\begin{align}
    \text{min} \quad \sum_{i=1}^N& (A_i y_i + B_i p_i + C_i p_i^2) \nonumber\\
&+ \lambda_1 (\sum_{i=1}^N p_i y_i - L)^2 \nonumber\\
&+ \lambda_2 [\sum_{i=1}^N (p_i - s_{i,1} - p_{i,min} y_i)^2] \nonumber\\
&+ \lambda_3 [\sum_{i=1}^N (p_i + s_{i,2} - p_{i,max} y_i)^2]
\label{eq:qubo_objective}
\end{align}

Where $s_{i, j}, p_i \in \mathbb{R}_{\ge 0}$, $y_i\in\set{0, 1}$ and empirically determined $\lambda_i$. We have now rewritten the the UC problem as a QUBO which will be useful for converting to a QAOA formulation so that all constraints are now contained in the objective. Since $y_i^2 = y_i$, the above problem reduces to a quadratic program by noting for fixed $p_i, s_{i, j}$ . 

Once the problem is reformulated, we use a SciPy classical minimizer to find the individual unit power values that minimize the cost function, as well as the optimal $\bm{\gamma}, \bm{\beta}$ variational parameters to use in QAOA. We use the minimize function from SciPy's optimize package \cite{scipyopt}, specifically using the Nelder-Mead algorithm \cite{scipynm} with the $\bm{\gamma}, \bm{\beta}, \bm{p}, \bm{s}$ as the arguments. 
For each iteration that the classical minimizer completes, we simulate a quantum processor with IBM Qiskit to execute QAOA \cite{qiskit}. This returns a probability distribution of the optimal combinations of units to be used, given the power values that the classical minimizer has designated to each unit. 

\subsection{Quantum Circuit}

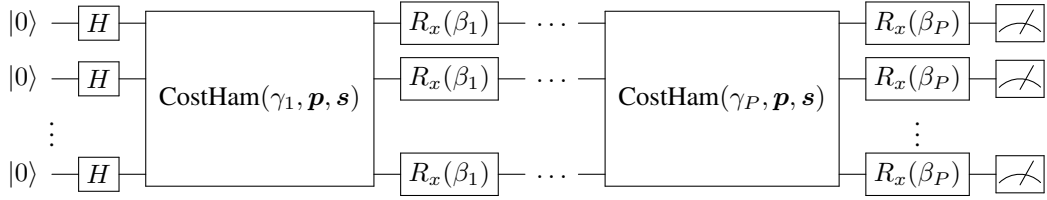
\begin{figure*}[h]
    \centering
$$
\Qcircuit @C=1em @R=0.5em {
\lstick{\ket{0}} & \gate{H} & \multigate{5}{\text{CostHam}(\gamma_1, \bm{p}, \bm{s})} & \gate{R_x(\beta_1)} & \qw & \dots & & \multigate{5}{\text{CostHam}(\gamma_P, \bm{p}, \bm{s})} & \gate{R_x(\beta_P)} & \meter \\
\lstick{\ket{0}} & \gate{H} & \ghost{\text{CostHam}(\gamma_1, \bm{p}, \bm{s})} & \gate{R_x(\beta_1)} & \qw & \dots & & \ghost{\text{CostHam}(\gamma_P, \bm{p}, \bm{s})} & \gate{R_x(\beta_P)} & \meter \\
\\
\vdots  &&&&&&&& \vdots \\
\\
\lstick{\ket{0}} & \gate{H} & \ghost{\text{CostHam}(\gamma_1, \bm{p}, \bm{s})} & \gate{R_x(\beta_1)} & \qw & \dots & & \ghost{\text{CostHam}(\gamma_P, \bm{p}, \bm{s})} & \gate{R_x(\beta_P)} & \meter \\
}
$$
\caption{Circuit diagram}
    \label{fig:circuit_diagram}
\end{figure*}

When the classical outer loop optimizer proposes $\bm{p}$ and $\bm{s}$, it induces a QUBO objective function, as described by Equation~\ref{eq:qubo_objective}. The classical optimizer is also responsible for proposing the $2P$ continuous variational parameters, $\bm{\gamma}, \bm{\beta}$. To comply with the standard form for QUBOs solved by QAOA, we convert from $y_i \in \{0, 1\}$ binary variables to $z_i \in \{+1, -1\}$ variables via the transformation $z_i = 2 y_i - 1$.

The actual QAOA circuit, given $\bm{\gamma}, \bm{\beta}$, is depicted in Fig.~\ref{fig:circuit_diagram}. Each Cost Hamiltonian step applies the operation with diagonal unitary matrix $e^{i \gamma_i H(\bm{\gamma}, \bm{\beta})}$ where $H(\bm{\gamma}, \bm{\beta})$ is a diagonal unitary matrix corresponding to Equation~\ref{eq:qubo_objective}. This step is the most expensive operation. Since the objective function is a QUBO, the sub-operations in the Cost Hamiltonian consist of a sequence of $e^{i \gamma_i Z \otimes Z}$ operations, where $Z$ is the Pauli-$Z$ matrix. We used the sympy library \cite{scipynm} to manage the construction of the QUBO and tracking of its coefficients. After the Cost Hamiltonian step, the mixing Hamiltonian is applied, which simply executes $R_x(\beta_i)$ on each qubit. These Cost and Mixing Hamiltonian steps are repeated $P$ times before terminating measurements on each qubit.

\section{Classical Baseline}
\begin{figure*}[h!]
    \centering
    \includegraphics[width=\textwidth]{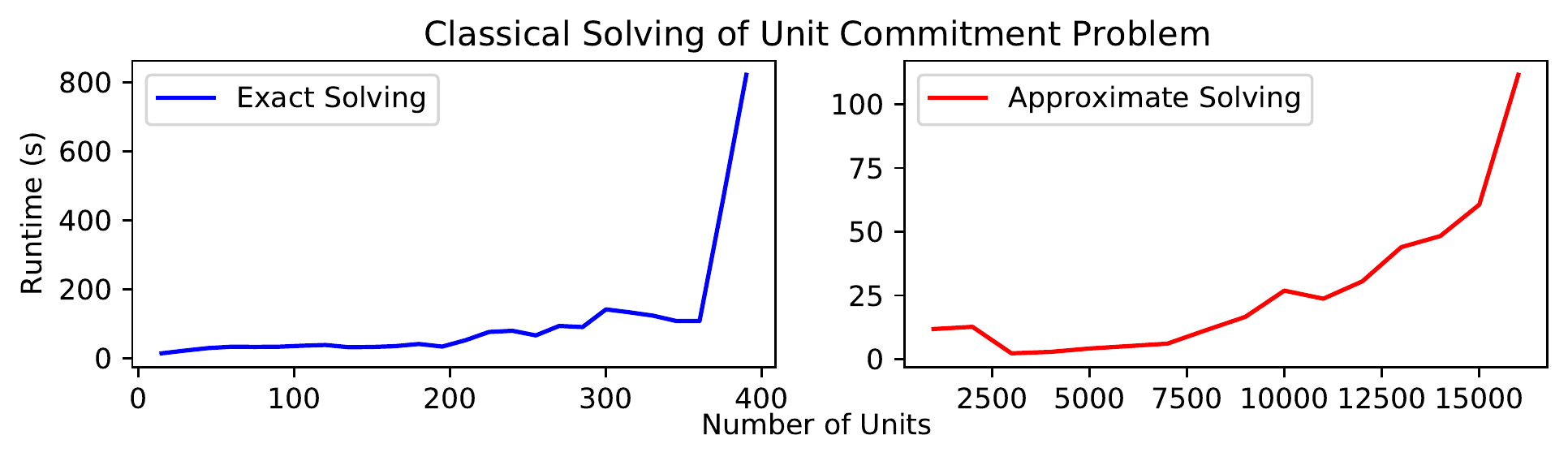}
    \caption{Classical solving of UC problems with IBM's CPLEX optimizer. For approximate solving, the algorithm finished with an objective function value within 8\% of the optimal solution.}
    \label{fig:classical_results}
\end{figure*}

To benchmark how existing classical approaches perform on the Unit Commitment problem, we solved UC instances of varying sizes both exactly and approximately using IBM's CPLEX Optimizer \cite{cplex}. Experiments were run using version 20.1 of IBM's CPLEX Optimizer on a single core of an Intel Core i7-8750H CPU. We used CPLEX to solve the UC instances in their MBO form. Fig.~\ref{fig:classical_results} shows results from solving these various UC problems. Approximate solving was done by searching for solutions with objective function value within 8\% of the optimal solution. 

CPLEX uses a branch-and-bound approach to solving the UC problems in MBO form, which although efficient in practice when compared to other classical methods, still has worst-case exponential scaling in runtime. The results show that, as expected, exact solving scales exponentially in runtime, quickly becoming intractable for large problem sizes. We also observe that approximate methods perform better in runtime compared to exact solving at the cost of solution quality, however still eventually see exponential scaling for larger problem sizes, as expected.

\section{Results}
To confirm our algorithm does approximate solutions to UC adequately, we ran different example systems and tracked the progress of numerous variables to ensure our algorithm was converging on approximate solutions each time it ran. One of our main examples was a real 10-unit system, specified with the parameters given in Table \ref{tab:10}. 
\begin{table*}[h]
    \begin{center}
    \begin{tabular}{|c|c|c|c|c|c|c|c|c|c|c|}
         \hline
         i & 1 & 2 & 3 & 4 & 5 & 6 & 7 & 8 & 9 & 10 \\
         \hline
         $p_{max, i}$ (MW) & 455 & 455 & 130 & 130 & 162 & 80 & 85 & 55 & 55 & 55 \\
         $p_{min, i}$ (MW) & 150 & 150 & 20 & 20 & 25 & 20 & 25 & 10 & 10 & 10 \\
         $A_i$ (\$) & 1000 & 970 & 700 & 680 & 450 & 370 & 480& 660 & 665 & 670 \\
         $B_i$ (\$/MW) & 16.19 & 17.26 & 16.60 & 16.50 & 19.70 & 22.26 & 27.74 & 25.92 & 27.27 & 27.79 \\
         $C_i$ (\$/$MW^2$) & .00048 & .00031 & .002 & .00211 & .00398 & .00712 & .0079 & .00413 & .00222 & .00173 \\
         \hline
    \end{tabular}
    \caption{The parameters of the 10-unit system}
    \label{tab:10}
    \end{center}
\end{table*}

\begin{figure*}[h!]
    \centering
    \includegraphics[width=\textwidth]{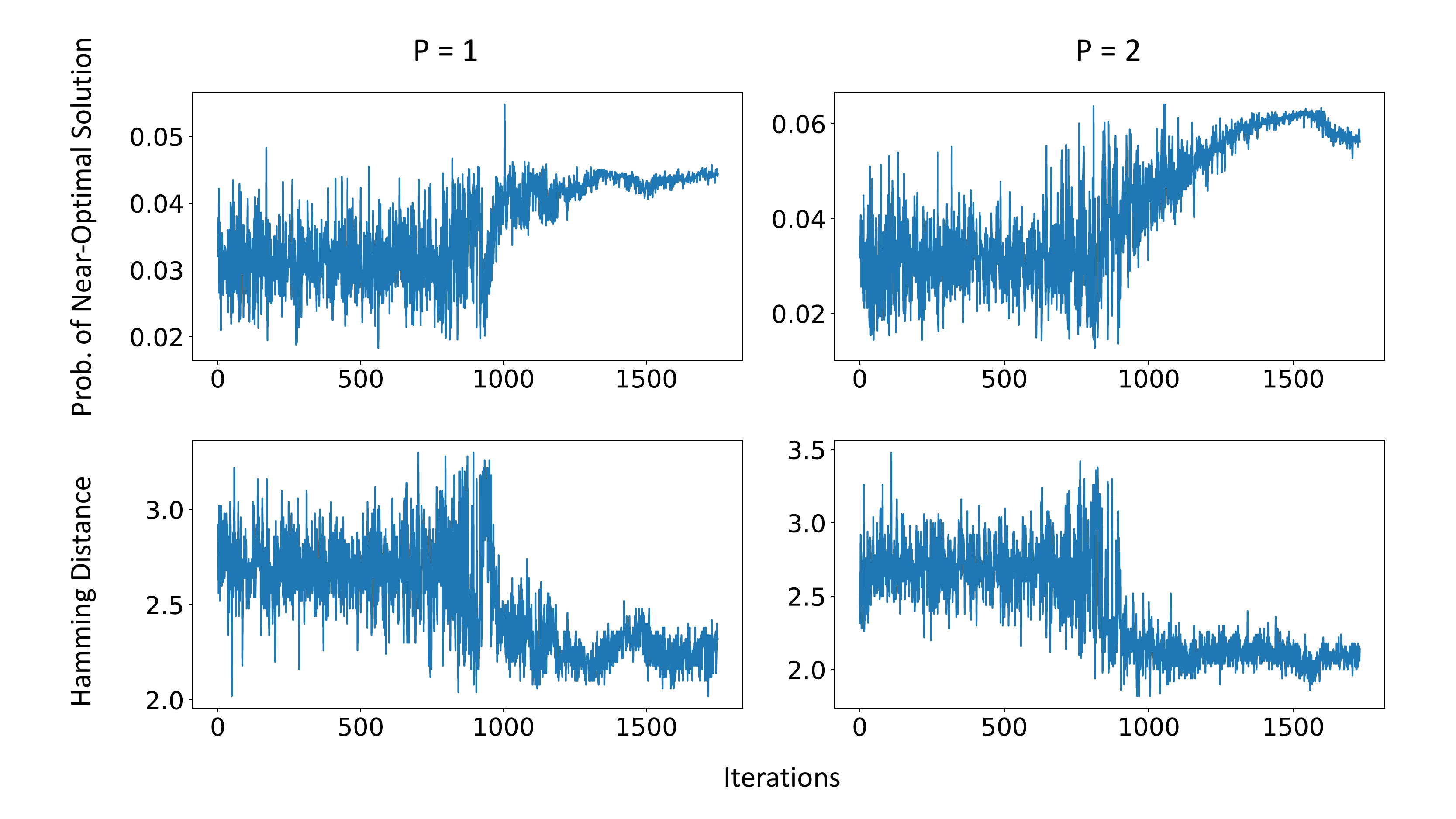}
    \caption{For a 10-unit system, both using $P=1$ (left) and $P=2$ (right), the total probability of finding a near optimal bitstring increases as our algorithm completes more iterations (top). Additionally, the average Hamming distance between each of the top 50 bitstrings and their closest near optimal solution decreases over time (bottom).}
    \label{fig:results_10}
\end{figure*}

Fig.~\ref{fig:results_10} demonstrates the results of our quantum approach on the 10-unit system in Table~\ref{tab:10}. The left and right plots corresponding to $P=1$ and $P=2$ depth for QAOA respectively. The top plots show the total probability of our algorithm returning a near-optimal solution \footnote{The set of near-optimal solutions were generated by classical brute force, with a the cutoff for ``near-optimal'' determined by best judgment. As a result, some systems may have many near optimal solutions, while others have only a few. The cutoff selection was done as fairly as possible, but note that these selections will affect the shape of the plot.}, i.e. an assignment of units to ON/OFF. Given a near-optimal bitstring, the actual power level assignments can be obtained by solving the induced quadratic program.

As the plots indicate, the probability of near-optimal solution increases over the course of variational optimization iterations in our hybrid approach. At iteration 0, the probability of achieving a near-optimal solution is essentially the same as random guessing. After 1500 iterations, the probability of near-optimal solution exceeds 4\% for $P=1$ and 6\% $P=2$. This both validates our general approach and demonstrates that increased performance is possible at higher $P$.

\begin{figure*}[h!]
    \centering
    \includegraphics[width=\textwidth]{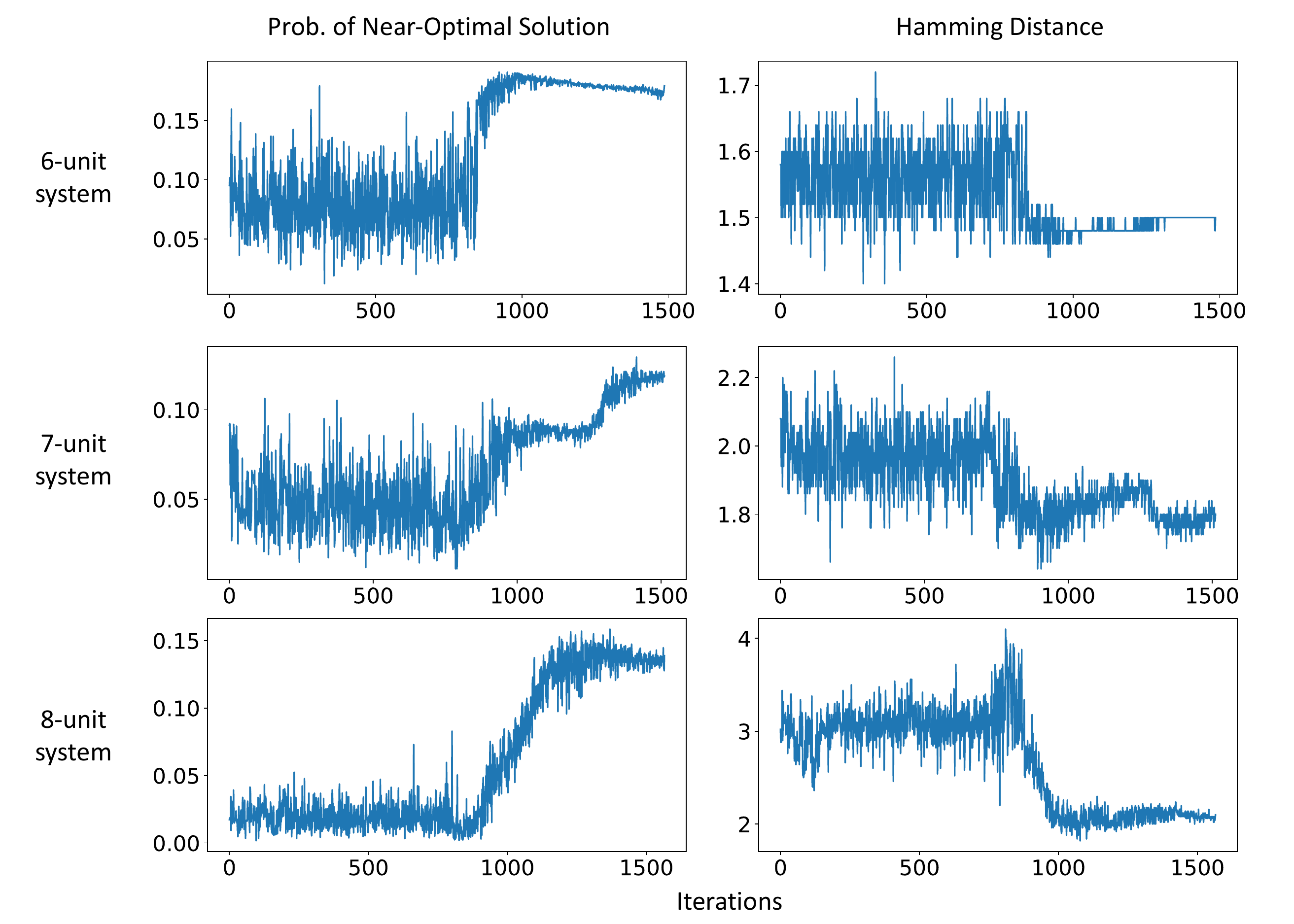}
    \caption{As in the case of the 10-unit system, when using $P=2$ steps of QAOA for a 6-, 7- and 8-unit system, the total probability of finding a near optimal bitstring increases over time (left), and the average Hamming distance between each of the top 50 bitstrings and their closest near optimal solution decreases over time (right).}
    \label{fig:results_other}
\end{figure*}

The bottom two plots show the average Hamming distance between each of the top 50 bitstrings returned by our algorithm and the near optimal solution that it is closest to. This metric decreases over the course of the variational optimization iterations, corroborating that our algorithm is converging to better solutions (i.e. ON/OFF assignments).

We also produced similar plots for sample 6-, 7-, and 8-unit systems. For each system, we used $P=2$ QAOA, and we found the same trends as for the 10-unit system, further supporting that our algorithm succeeds in driving towards a near-optimal solution (assignment of units to ON/OFF). For brevity, we leave out details of the parameters for these systems; the plot of results are in Fig.~\ref{fig:results_other}.

\section{Conclusion}
We have introduced a hybrid quantum-classical approach to Unit Commitment. Our approach uses QAOA to turn a QUBO instance into a continuous optimization problem over variational parameters $\bm{\gamma}, \bm{\beta}$. A classical optimizer is then able to simultaneously optimize over these continuous parameters as well as the power assignments for each unit.

Figures~\ref{fig:results_10} and \ref{fig:results_other} validate the correctness and promise our approach. As demonstrated for 6-, 7-, and 8-, and 10- unit sytems (which entails quantum simulation of a circuit with equal number of qubits), our variational approach boosts the probability of finding a near-optimal bitstring (i.e. assignment of units to ON/OFF) far beyond naive brute force or random guessing. Moreover, increasing the QAOA depth parameter, $P$, improves performance.

For smaller systems with only a few hundred units, existing classical solvers are able to perform well. However, as our classical simulation results indicate, these classical solvers are limited for larger systems, because their runtime scales exponentially. By contrast, our hybrid quantum-classical algorithm could maintain effectiveness at larger system sizes, since each QAOA circuit has logical depth of just $O(N^2P)$.

In the future, it will be important to run our algorithm on a real machine rather than simulating the quantum circuits. We anticipate that real evaluation will pose a variety of challenges not seen in our ideal classical simulation. For example, superconducting qubits have sparse connectivity, which could incur a large SWAP overhead (though this could be mitigated with techniques like swap networks \cite{kivlichan2018quantum, tomesh2020coreset}). In addition, the recently discovered phenomenon of noise-induced barren plateaus \cite{wang2020noise} could hinder successful execution in near-term quantum computers. However, we expect quantum error correction to emerge in hardware over the next decade; in this medium-term era, noise would no longer be an obstacle.
\clearpage

\section*{Acknowledgement}

This work is funded in part by EPiQC, an NSF Expedition in Computing, under grants CCF1730082/1730449; in part by STAQ under grant NSF Phy-1818914; in part by DOE grants DE-SC0020289 and DE-SC0020331; and in part by NSF OMA2016136 and the Q-NEXT DOE NQI Center. This material is based upon work supported by the National Science Foundation under Grant No. 2110860.

Disclosure: Fred Chong is Chief Scientist at Super.tech and an advisor to Quantum Circuits, Inc.
\printbibliography
\end{document}